%Paper: hep-ph/9405250
%From: Jean-Bruno Erismann <jb@cptsu5.univ-mrs.fr>
%Date: Mon, 9 May 94 10:15:44 +0200

\magnification=1200
\catcode `@=11
\hsize 14truecm
\vsize 24truecm
\hoffset=0,9truecm

\def\build#1_#2^#3{\mathrel{\mathop{\kern 0pt#1}\limits_{#2}^{#3}}}

\def\m{\medskip}
\def\b{\bigskip}
\def\n{\noindent}
\def\o{\overline}

\font\eightrm=cmr8

\font\ten=cmbx10 at 13pt
\font\twelve=cmbx10
\font\boldit=cmbxti10

\baselineskip 15pt
{

\centerline{\twelve Centre de Physique Th\'eorique - CNRS - Luminy,
Case 907}
\centerline{\twelve F--13288 Marseille Cedex 9 - France }
\centerline{\bf Unit\'e Propre de Recherche 7061}

\vskip 2truecm

\centerline{\ten PARTON DISTRIBUTIONS FROM {\boldit W}$^{\pm}$ AND
{\boldit Z}}
\centerline{\ten PRODUCTION IN POLARIZED {\boldit pp} AND {\boldit
pn}}
\centerline{\ten COLLISIONS AT
RHIC}

\bigskip

\centerline{\bf Claude BOURRELY and Jacques SOFFER}

\vskip 2truecm

\centerline{\bf Abstract}

\medskip

We study the production of $W^{\pm}$ and $Z$ gauge bosons in
proton-proton and proton-neutron collisions up to a center-of-mass
energy $\sqrt s=0.5\ TeV$, in connection with the realistic
possibility of having proton-deuteron collisions at RHIC with a
high luminosity. We stress the importance of measuring unpolarized
cross sections for a better determination of the flavor asymmetry
of the sea quarks. Since polarized proton beams will be available at
RHIC, we evaluate several spin-dependent observables like, double
helicity asymmetries $A_{LL}$ as a test of the sea quark
polarization, double transverse spin asymmetries $A_{TT}$ as a
practical way to determine  a new structure function, the nucleon
transversity distribution $h_1(x)$, and parity-violating
asymmetries $A_L$ as a further test of our present knowledge of
polarized parton distributions.

\vskip 3truecm

\noindent Number of figures : 19

\bigskip

\noindent January 1994

\noindent CPT-94/P.3000

\bigskip

\noindent anonymous ftp or gopher : cpt.univ-mrs.fr

\footline={}
\vfill\eject}
\pageno=1

\baselineskip 18pt

\item{\hbox to\parindent{\enskip {\bf 1.}\hfill}} {\bf INTRODUCTION}

\m

A Relativistic Heavy Ion Collider (RHIC) is now under construction
at Brookhaven National Laboratory and, already more than two years
ago, it was realized that one should propose a very challenging
physics programme$^{[1]}$, provided this machine could be ever used
as a polarized $pp$ collider. Of cour\-se all these considerations
relie on the foreseen key parameters of this new facility, i.e. a
luminosity of $2.10^{32}cm^{-2}sec^{-1}$ and an energy of $50-250\
GeV$ per beam with a polarization of about $70$\%. Since then, the
RHIC Spin Collaboration (RSC) has produced a letter of intent$^{[2]}$
and has undertaken several serious studies in various areas which
have led to a proposal$^{[3]}$ which has now been fully approved.

Among the very many basic hadronic reactions which will be studied
in this experimental programme, we recall that jet production and
direct photon production are sensitive to the size of the gluon
polarization, whereas Drell-Yan lepton pair  production will allow to
determine the magnitude and the sign of the sea quark
polarization$^{[1,4]}$. Recent measurements of proton and neutron
$F_2$ structure functions by the NMC at CERN$^{[5]}$ have yielded a
violation of the Gottfried sum rule$^{[6]}$ whose most natural
interpretation is the evidence for a flavor asymmetry in the light
sea quarks (i.e. $\o d(x)>\o u(x)$). In a recent paper$^{[7]}$, we
have shown that this can be studied further in the production of
$W^{\pm}$ and $Z$ bosons in a proton-proton collider because it is
dominated by quark-antiquark annihilation and therefore sensitive to
the sea distributions. It was stressed that in the central region the
predicted unpolarized cross sections can vary by a large factor
depending on, whether or not, one assumes a flavor symmetric sea. The
measurement of charge asymmetries in the $W^+$ and $W^-$ production
was also suggested as an interesting possibility$^{[8]}$. Parity
violating asymmetries with either one beam polarized $A_L$ or with
two beams polarized $A_{LL}^{PV}$ have been calculated$^{[7]}$ and we
showed that their measurement, will give us a good calibration of the
polarization of quarks and antiquarks, for both $u$ and $d$. In this
paper we are further investigating $W^{\pm}$ and $Z$ bosons
production and, in particular, we emphasize some new possibilities if
one can measure simultaneously proton-proton and proton-neutron
collisions, assuming the feasibility of tagging the spectator in
proton-deuteron interactions$^{[9]}$. We will also consider several
spin dependent observables, with longitudinally and transversely
polarized proton beams, which contain useful information on the
polarized parton distributions.

The paper is organized as follows. In section 2 we compare
unpolarized cross sections in $pp$ and $pn$ collisions and we show it
allows a much improved determination of flavor asymmetry of the sea
quarks. In section 3 we evaluate double helicity asymmetries $A_{LL}$
for $pp$ collisions with both proton beams polarized and parity
violating asymmetries $A_L$ for $pn$ collisions with only one proton
beam longitudinally polarized. Finally in section 4 we consider the
situation of $pp$ collisions with both proton beams transversely
polarized and we will see that only in the case of $Z$ production,
one can extract the new structure function $h_1(x)$ $^{[10]}$, the
{\it transversity} distribution, which indeed measures the
correlation between the left-handed and the right-handed quarks in a
transversely polarized proton. We will give our concluding remarks in
section 5.

\b

\item{\hbox to\parindent{\enskip {\bf 2.}\hfill}} {\bf UNPOLARIZED
CROSS SECTIONS}

\m

Let us recall that the differential cross section for the reaction
$$pp\to W^{\pm}+\hbox{ anything}\eqno (1)$$
can be computed directly in the Drell-Yan picture in terms of the
{\it dominant} quark-antiquark fusion reactions $u\o d\to W^+$ and
$\o ud\to W^-$ and we have for the Standard Model $W^+$ production
$${d\sigma^{W^+}_{pp}\over dy}=G_F\pi\sqrt 2\tau\cdot{1\over
3}\left[u(x_a,M^2_W)\o d(x_b,M^2_W)+(u\leftrightarrow \o d)
\right]\eqno (2)$$
with
$$G_F={\pi\alpha\over\sqrt 2M^2_W\sin^2\theta_W}\ \raise 2pt\hbox{,}\
x_a=\sqrt{\tau}e^y\ ,\ x_b=\sqrt{\tau}e^{-y}\ ,\ \hbox{and}\
\tau=M^2_W/s\ .$$

Here $\sin^2\theta_W$ is the weak mixing angle and quark flavors are
interchanged in eq.(2) for $W^-$ production. Clearly these $y$
distributions are symmetric under $y\to -y$. For the reaction
$$pn\to W^{\pm} +\ \hbox{anything}\eqno (3)$$
under the assumption of isospin invariance for the nucleon, i.e.
$u_p(x)=d_n(x)=u(x)$, etc... (also for antiquarks), one has for $W^+$
production
$${d\sigma^{W^+}_{pn}\over dy}=G_F\pi\sqrt 2\tau\cdot{1\over
3}\left[u(x_a,M^2_W)\o u(x_b,M^2_W)+\o d(x_a,M^2_W)d(x_b,M^2_W)
\right]\eqno (4)$$
which is no longer symmetric under $y\to -y$ and it simply follows
that for $W^-$ production one has
$${d\sigma^{W^-}_{pn}\over
dy}(y)={d\sigma^{W^+}_{pn}\over dy}(-y)\ \cdotp\eqno(5)$$

All these cross sections will strongly be dependent on the antiquark
distributions whose present determination is more uncertain than that
of $u(x)$ and $d(x)$. Following ref.[7] we will take the valence
quarks $u_v(x,Q^2)$ and $d_v(x,Q^2)$ from ref.[11] with the
appropriate evolution from $Q^2=4$ GeV$^2$ to $Q^2=M^2_W$. We will
assume that the antiquark and sea quark distributions go like
$(1-x)^{10}$, flavor symmetry breaking being obtained by taking
$$x\delta\o q(x)=x\o u(x)-x\o d(x)=ax^b(1-x)^{10}\eqno (6)$$
with $a=-0.22$ and $b=0.45$. For our calculations we have used
$M_W=80.22\ GeV$, the latest value given in the Particle
Properties Data Booklet. This value is lower than the one used
in ref.[7] and it leads to slightly larger predictions for $pp$
collisions. The results are shown in figs. 1 and 2 for two different
center of mass energies
$\sqrt s=350$ and $500\ GeV$, the lower one being the maximum energy
in $pn$ collisions when the proton and the deuteron beam momenta are
$250\ GeV/c$. Given the luminosity mentioned above, we see that one
expects a fairly large event rate for all these cases. As in ref.[7]
we could have studied the sensitivity of these cross sections to
various choices of the antiquark distributions, but in order to avoid
some uncertainties related to their absolute normalization and to pin
down the flavor asymmetry of the sea we introduce the following
quantity
$$R_W={d\sigma^{W^+}_{pp}/dy - d\sigma^{W^+}_{pn}/dy
+d\sigma^{W^-}_{pp}/dy - d\sigma^{W^-}_{pn}/dy\over
d\sigma^{W^+}_{pp}/dy + d\sigma^{W^+}_{pn}/dy +d\sigma^{W^-}_{pp}/dy
+ d\sigma^{W^-}_{pn}/dy}\ \cdotp\eqno (7)$$

In terms of parton distributions, it simply reads
$$R_W=-{\delta q(x_a)\delta\o q(x_b)+(x_a\leftrightarrow x_b)\over
\sigma q(x_a)\sigma\o q(x_b)+(x_a\leftrightarrow x_b)}\eqno (8)$$
where $\displaystyle\left({\sigma\atop\delta}\right) q(x)=u(x)\pm
d(x)$ and similarly for antiquarks. Clearly $R_W$ is symmetric under
$y\to -y$ and $R_W=0$ if the sea is flavor symmetric, i.e. $\o
u(x)=\o d(x)$ that is $\delta\o q(x)\equiv 0$. The result of the
calculation for our choice of the flavor symmetry breaking (see
eq.(6)) is shown in fig.3 for two different energies and it provides
a very clear test for the antiquark distributions. We find that $R_W$
is positive which is due to the fact that for our choice $\delta\o
q(x)<0$, because $\o d(x)>\o u(x)$.

In the case of $Z$ production, similar calculations can be done in
the Drell-Yan picture in terms of $u\o u$ and $d\o d$ annihilations.
For $pp$ collisions we have $${d\sigma^Z_{pp}\over
dy}={G_F\pi\tau\over \sqrt 2}\cdot{1\over 3}\sum_{i=u,d}^{}
(a^2_i+b^2_i)\left[q_i(x_a,M^2_Z)\o
q_i(x_b,M^2_Z)+(x_a\leftrightarrow x_b)\right]\eqno (9)$$
where $a_i$
and $b_i$ denote the vector and axial vector coupling  constants of
the quark $i$ to the $Z$ boson. For $pn$ collisions one obtains the
corresponding expression $d\sigma^Z_{pn}/dy$ by making the
appropriate substitutions as above for the case of $W^{\pm}$
production. The results are shown in figs.4 and 5 for $M_Z=91.173\
GeV$ at two different energies where we see, as expected, a symmetric
distribution for $pp$ and asymmetric for $pn$ collisions.

Let us also consider a quantity which is less sensitive to absolute
normalization uncertainties i.e.
$$R_Z={d\sigma^{Z}_{pp}/dy - d\sigma^{Z}_{pn}/dy\over d
\sigma^{Z}_{pp}/dy + d\sigma^{Z}_{pn}/dy}\eqno(10)$$
and which reads in terms of the parton distributions taken at
$Q^2=M^2_Z$
$$\eqalignno{R_Z=&\left\{(a^2_u+b^2_u)\left[u(x_a)\delta\o q(x_b)+\o
u(x_a)\delta q(x_b)\right]\right.\cr &\left. - (a^2_d+b^2_d)
\left[d(x_a)\delta\o q(x_b)+\o d(x_a)\delta q(x_b)\right]\right\}/\cr
&\sum_{i=u,d}^{}(a^2_i+b^2_i) \left[q_i(x_a)\sigma\o q(x_b)+\o
q_i(x_a)\sigma q(x_b)\right]&(11)\cr}$$

In this case, even for $\delta\o q\equiv 0$ we don't have $R_Z=0$,
because the couplings of $u$ and $d$ quarks are different and we find
$R_Z$ negative because
$\left(a_d^2+b_d^2\right)>\left(a_u^2+b_u^2\right)$. The results
are given in figs. 6a and b for two different energies where the
dashed curves correspond to $\delta\o q\equiv 0$ and solid
curves to our choice of the flavor symmetry breaking (see eq. (6)).
Although the $Z$ production will be less copious this comparison
between $pp$ and $pn$ collisions will also allow to discriminate
between $\delta\o q\equiv 0$ and $\delta\o q\ne 0$.

\b

\item{\hbox to\parindent{\enskip {\bf 3.}\hfill}} {\bf HELICITY
ASYMMETRIES WITH LONGITUDINALLY\hfill\break POLARIZED
PROTON BEAMS}

\m

Since RHIC will be built to be used as a polarized $pp$ collider,
let us now investigate what we can learn from the measurement
of various spin-dependent observables. We first start with one
longitudinally polarized proton beam and we consider the
parity-violating helicity asymmetry defined as
$$A_L={d\sigma_- -d\sigma_+\over d\sigma_-+ d\sigma_+}\
\cdotp\eqno(12)$$
Here $\sigma_h$ denotes the cross section where the initial
proton has helicity $h$. This asymmetry will be expressed in
terms of the parton helicity asymmetries $\Delta
f(x,Q^2)=f_+(x,Q^2)-f_-(x,Q^2)$ where $f_{\pm}(x,Q^2)$ denote the
parton distributions in a polarized nucleon either with helicity
parallel $(+)$ or antiparallel $(-)$ to the parent nucleon helicity.
In the Standard Model the $W$ is a purely left handed current and
this asymmetry reads simply, for $W^+$ production in $pp$
collisions,
$$A_L(y)={\Delta u\left(x_a,M_W^2\right)\o
d\left(x_b,M_W^2\right)-\left(u\leftrightarrow\o d\right)\over u
\left(x_a,M_W^2\right)\o d\left(x_b,M_W^2\right)+
\left(u\leftrightarrow\o d\right)}\eqno(13)$$
assuming the proton $a$ is polarized. For $W^-$ production quark
flavors are interchanged. This asymmetry, together with two
other parity-violating helicity asymmetries related to it, have
been studied in ref.~[7], so here we will rather consider the case
of $pn$ collisions which will be also combined with some of our
previous results on $pp$ collisions from ref.~[7]. For $W^+$
production in $pn$ collisions one has
$$A_L(y)={\Delta u\left(x_a,M_W^2\right)\o
u\left(x_b,M_W^2\right)- \Delta\o d\left(x_a,M_W^2\right)
d\left(x_b,M_W^2\right)\over u
\left(x_a,M_W^2\right)\o
u\left(x_b,M_W^2\right)+\o d\left(x_a,M_W^2\right)
d\left(x_b,M_W^2\right)}\eqno(14)$$
assuming the proton is polarized\footnote{$^{(1)}$}{\eightrm
In the proton-deuteron collisions we assume that only the proton
beam is polarized,}\footnote{}{\eightrm  because it is not clear to
us that it will be possible to have polarized deuteron beams.}. In
order to calculate these asymmetries we have to use a model for
the various partons (valence quark, sea quark, antiquark) helicity
asymmetries
$\Delta u_v,\ \Delta u_s,\ \Delta \o u,\ \Delta d_v,\ \Delta d_s,\
\Delta \o d$ compatible with polarized deep inelastic
scattering data. We recall that in leptoproduction with both lepton and nucleon
longitudinally polarized, one extracts the structure function
$g_1(x)=1/2\displaystyle\sum_i e_i^2\left[\Delta q_i(x)+\Delta
\o q_i(x)\right]$ which is, so far, our unique source of
information on the quark and antiquark helicity asymmetries. Many
parametrizations have been proposed in the litterature, but following
ref.~[7] we will adopt a recent suggestion$^{[12]}$ to construct
polarized quark distributions from unpolarized quark distributions,
based on the extensive use of the Pauli exclusion principle. In this
approach\footnote{$^{(2)}$}{\eightrm For a more recent work
on the description of quark distributions in terms of
Fermi-}\footnote{}{\eightrm Dirac distributions see also ref.~[13].}
one successfully relates the violation of the Gottfried sum rule to
the Ellis-Jaffe sum rule$^{[14]}$ defect reported earlier by the EMC
in the proton case$^{[15]}$. So concerning the valence distributions
we will take
$$\Delta u_v(x)=u_v(x)-d_v(x)\quad\hbox{and}\quad\Delta
d_v(x)=-1/4\ d_v(x),\eqno(15)$$
whereas for sea quark, and antiquark, assuming that $\Delta
q_s(x)=\Delta\o q_s(x)$, for the $\o u$'s we will take following
ref.~[7]
$$\Delta\o u(x)=\o u(x)-\o d(x)\equiv\delta\o q(x)\eqno(16a)$$
which leads to a large negative $\o u$ polarization. Concerning
the $\o d$'s, in ref.~[7] their polarization was taken to be small
and positive, but our analysis$^{[13]}$ of the recent SLAC data on
polarized neutron deep inelastic scattering$^{[16]}$ is consistent
with
$$\Delta\o d(x)\equiv 0,\eqno(16b)$$
and this will be assumed from now on. We show in fig.~7 our
choice of the various parton polarizations $\Delta q/q$ taken at
$Q^2=M_W^2$. Returning now to $A_L$, we show the results of our
calculations for $pn$ collisions in figs~8a,~b at two different
energies and we observe that there is very little energy dependence.
The general trend of $A_L$ is similar to what we obtained in
ref.~[7] for $pp$ collisions. It can be understood from eq.~(14)
and we see that for $y=-1,\ A_L$ is sensitive to the antiquark
polarizations since
$$A_L^{W^+}\sim -{\Delta\o d\over \o d}\quad\hbox{and}\quad
A_L^{W^-}\sim -{\Delta\o u\over \o u}\eqno(17)$$
whereas for $y=+1$, it is sensitive to the quark polarizations
since
$$A_L^{W^+}\sim{\Delta u\over u}\quad\hbox{and}\quad
A_L^{W^-}\sim{\Delta d\over d}\ \cdotp\eqno(18)$$
So it is easy from fig.~7 to anticipate the results shown in
figs.~8a, b and in particular if $\Delta\o d$ is not equal to zero,
it will clearly show up in $A_L^{W^+}$ near $y=-1$. From the
measurement of the cross section $d\sigma/dy$ and the
parity-violating asymmetry $A_L$ for the four cases $pp\to
W^{\pm}$ and $pn\to W^{\pm}$ one can obtain a quantity which is
less sensitive to absolute normalization uncertainties, that is
$$a_L^W={(A_L d\sigma/dy)_{pp}^{W^+}-(A_L
d\sigma/dy)_{pn}^{W^+}+(A_L d\sigma/dy)_{pp}^{W^-}-(A_L
d\sigma/dy)_{pn}^{W^-}
\over d\sigma_{pp}^{W^+}/dy+d\sigma_{pn}^{W^+}/dy
+d\sigma_{pp}^{W^-}/dy+d\sigma_{pn}^{W^-}/dy}\
\cdotp\eqno(19)$$
In terms of parton distributions taken at $Q^2=M_W^2$ it simply
reads
$$a_L^W=-{\sigma\Delta q(x_a)\delta\o q(x_b)+\sigma\Delta
\o q(x_a)\delta q(x_b)\over\sigma q(x_a)\sigma\o q(x_b)+
(x_a\leftrightarrow x_b)}\eqno(20)$$
where $\sigma\Delta q(x)=\Delta u(x)+\Delta d(x)$ and similarly
for antiquarks. So in a si\-tu\-ation where the sea is flavor
symmetric i.e. $\delta\o q(x)\equiv 0$ {\it and} the antiquarks
are unpolarized i.e. $\Delta\o q(x)\equiv 0$, one has $a_L^W\equiv
0$. Our results are shown in figs.~9a, b for two different
energies. The solid curve corresponds to $\delta\o q(x)\ne 0$ and
$\Delta\o q(x)\ne 0$ so $a_L^W$ is positive and lies between
5\% and 15\%. However if we assume flavor symmetry breaking
for the sea i.e. $\delta\o q(x)\ne 0$, but unpolarized antiquarks
$\Delta\o q(x)=0$, we get the small dashed curve, which
compared to the solid curve exhibits clearly the sensitivity to
$\Delta\o q(x)$ near $y=-1$. Finally one can also consider the
case where $\delta\o q(x)=0$ so $\Delta\o u(x)=0$ and $\Delta\o
d(x)\ne 0$ at variance with eq.~(16b), then we would have
$a_L^W\ne 0$ and of opposite sign to $\Delta\o d(x)$.

We now turn to the case of $Z$ production where similar
calculations can be done. For $A_L$ we obtain the following
expression
$$A_L={\left[2a_u b_u\Delta u(x_a)\o d(x_b)+2a_d b_d\Delta
d(x_a)\o u(x_b)\right]-(q\leftrightarrow\o q)
\over
\left[\left(a_u^2+b_u^2\right) u(x_a)\o
d(x_b)+\left(a_d^2+b_d^2\right) d(x_a)\o
u(x_b)\right]+(q\leftrightarrow\o q)}\ \cdotp\eqno(21)$$
The results are shown in fig.~10 at two different energies. We
get large positive values near $y=+1$, whereas it is almost zero
near $y=-1$, at variance with the case of $pp$ collisions where it
was also large and positive near $y=-1$$^{[7]}$ and this is related to
the fact that $\o d(x)>\o u(x)$ as we will see now. Like for
$W^{\pm}$ production combining $pp$ and $pn$ measurements, we
now consider the following quantity
$$a_L^Z={\left(A_L d\sigma/dy\right)_{pp}^Z-\left(A_L
d\sigma/dy\right)_{pn}^Z
\over d\sigma_{pp}^Z/dy + d\sigma_{pn}^Z/dy}\ \cdotp\eqno(22)$$
In terms of the parton distributions taken at $Q^2=M_Z^2$ it
simply reads
$$a_L^Z={\left[2a_u b_u\Delta u(x_a)-2a_d b_d\Delta
d(x_a)\right]\left[\o u(x_b)-\o d(x_b)\right]-(q\leftrightarrow\o
q)\over
\left[\left(a_u^2+b_u^2\right)u(x_a)+ \left(a_d^2+b_d^2\right)
d(x_a)\right]\left[\o u(x_b)+\o d(x_b)\right]+(q\leftrightarrow\o
q)}\ \cdotp\eqno(23)$$
Clearly in a situation where the sea is flavor symmetric {\it and}
the antiquarks are unpolarized one has $a_L^Z\equiv 0$. We show
our results in figs~11a, b for two different energies. The solid
curve corresponds to $\delta\o q(x)\ne 0$ and $\Delta\o q(x)\ne
0$ so $a_L^Z$ is around +10\% near $y=-1$ and reflects the
trend of $A_L$ in $pp$ collisions whereas $a_L^Z$  is around
-10\% near $y=+1$ and is driven by $A_L$ in $pn$ collisions. For
unpolarized antiquarks $\Delta\o q(x)=0$, we get the small
dashed curve which goes to zero near $y=-1$ as $A_L$ in $pp$
collisions for this case, but it follows the solid curve near $y=+1$.
Finally we can also consider the case where $\delta\o q(x)=0$, so
$\Delta\o u(x)=0$ and $\Delta\o d(x)\ne 0$~; then we would have
$a_L^Z\ne 0$ and with the sign of $\Delta\o d(x)$.

Before closing this section let us consider, in $pp$ collisions
where both proton beams are polarized, another observable which
is very sensitive to antiquark polarizations, that is the
parity-conserving double helicity asymmetry $A_{LL}$ defined as
$$A_{LL}={d\sigma_{++}+d\sigma_{--}-
d\sigma_{+-}-d\sigma_{-+}
\over d\sigma_{++}+d\sigma_{--}+
d\sigma_{+-}+d\sigma_{-+}}\ \cdotp\eqno(24)$$
Here $\sigma_{h_1 h_2}$ is the cross section where the initial
protons have helicities $h_1$ and $h_2$. This asymmetry, in
$W^+$ production, reads simply
$$A_{LL}(y)=-{\Delta u\left(x_a,\ M_W^2\right)\Delta\o
d\left(x_b,\ M_W^2\right)+(u\leftrightarrow\o d)\over
u\left(x_a,\ M_W^2\right)\o
d\left(x_b,\ M_W^2\right)+(u\leftrightarrow\o d)}\
\cdotp\eqno(25)$$
For $W^-$ production quark flavors are
interchanged. It is clear that $A_{LL}(y)=A_{LL}(-y)$ and that
$A_{LL}\equiv 0$ if the antiquarks are not polarized, i.e. $\Delta\o
u(x)=\Delta\o d(x)\equiv 0$. Similarly for $Z$ production we find
$$A_{LL}(y)=-{\displaystyle\sum_{i=u,d}\left(a_i^2+b_i^2\right)
\left[\Delta q_i\left(x_a,\ M_Z^2\right) \Delta\o q_i \left(x_b,\
M_Z^2\right)+ (x_a\leftrightarrow x_b)\right]\over
\displaystyle\sum_{i=u,d}\left(a_i^2+b_i^2\right)\left[q_i
\left(x_a,\ M_Z^2\right) \o q_i \left(x_b,\ M_Z^2\right)+
(x_a\leftrightarrow x_b)\right]}\eqno(26)$$
which will vanish for unpolarized antiquarks. We show in
figs~12a, b our predictions for the three cases at two different
energies and we now try to understand these results. Clearly as a
consequence of eq.~(16b) $A_{LL}\equiv 0$ for $W^+$ production
but if $\Delta\o d(x)\ne 0$, it would be non-zero and of opposite
sign to $\Delta\o d(x)$. For $W^-$ production for $y=0$ we get
$A_{LL}=-{\Delta d\over d}{\Delta\o u\over\o u}$ evaluated at
$x=M_W/\sqrt s$ which gives around +10\% according to fig.~7.
{}From the trend of the $d$ and $\o u$ polarizations shown in fig.~7
we also expect $A_{LL}$ to be almost constant for $-1<y<+1$.
Finally for $Z$ production as a consequence of eq.~(16b) $A_{LL}$
does not depend on the $d$ quark polarization. For $y\simeq\pm
1$ one has approximately $A_{LL}\sim -{\Delta u(x_a)\over
u(x_a)}{\Delta\o u(x_b)\over\o u(x_b)}$ evaluated at $x_a\sim
0.65$ and $x_b\sim 0.10$ for $\sqrt s=350\ GeV$ that is,
according to fig.~7, $A_{LL}\sim+40$\%. For $\sqrt s=500\ GeV$
we get $A_{LL}\sim +25$\%. The values at $y=0$ are smaller
because of the maximum of the unpolarized cross section.

\b

\item{\hbox to\parindent{\enskip {\bf 4.}\hfill}} {\bf DOUBLE
SPIN TRANSVERSE ASYMMETRY WITH\hfill\break POLARIZED PROTON BEAMS}

\m

So far we have considered collisions involving only longitudinally
pola\-ri\-zed proton beams, but of course at RHIC transversely
polarized protons will be available as well$^{[3]}$. This new
possibility is extremely appealing because of recent progress in
understan\-ding transverse spin effects in QCD, both at leading
twist$^{[10]}$ and higher twist levels$^{[17]}$. For the case of the
nucleon's helicity, its distribution among the various quarks and
antiquarks can be obtained in polarized deep inelastic scattering
from the measurement of the structure function $g_1(x)$
mentioned above. However this is not possible for the {\it
transversity} distribution $h_1(x)$ which describes the state of
a quark (antiquark) in a transversely polarized nucleon. The
reason is that $h_1(x)$, which measures the correlation between
right-handed and left-handed quarks, decouples from deep
inelastic scattering. Indeed like $g_1(x),\ h_1(x)$ is leading
- twist and it can be measured in Drell-Yan lepton-pair production
with both initial proton beams transversely polarized$^{[10]}$. Other
possibilities have been suggested$^{[18]}$ but in the framework of
this paper, we will propose  a practical way to determine
$h_1(x)$ using gauge boson production in $pp$ collisions with
protons transversely polarized. Let us consider the double spin
transverse asymmetry defined as
$$A_{TT}={d\sigma_{\uparrow\uparrow}-
d\sigma_{\uparrow\downarrow}
\over d\sigma_{\uparrow\uparrow}+
d\sigma_{\uparrow\downarrow}}\eqno(27)$$
where $\sigma_{\uparrow\uparrow}
(\sigma_{\uparrow\downarrow})$ denotes the cross section with
the two initial protons transversely polari\-zed in the same
(opposite) direction. Assuming that the underlying parton
subprocess is quark-antiquark annihilation, we easily find for $Z$
production
$$A_{TT}={\displaystyle\sum_{i=u,d}\left(b_i^2-a_i^2\right)
\left[ h_1^{q_i}(x_a) h_1^{\o q_i}(x_b)+(x_a\leftrightarrow
x_b)\right]\over
\displaystyle\sum_{i=u,d}\left(a_i^2+b_i^2\right)
\left[ q_i(x_a)\o q_i(x_b)+(x_a\leftrightarrow
x_b)\right]}\ \cdotp\eqno(28)$$
This result generalizes the case of lepton-pair production$^{[10]}$
through an off shell photon $\gamma^{\star}$ and corresponding
to $b_i=0$ and $a_i=e_i$, the electric charge of $q_i$. For
$W^{\pm}$ production, which is pure left-handed and therefore
does not allow right-left interference, we expect $A_{TT}=0$,
since in this case $a_i^2=b_i^2$. This result is worth checking
experimentally.

We recall that $h_1(x)$ has never been measured, so to evaluate
$A_{TT}$ one has to make an assumption on the possible magnitude
of $h_1(x)$ for quarks and antiquarks. According to the MIT bag
model$^{[10]}$ one expects relativistic effects to increase the
magnitude of $h_1(x)$ and since in the non relativistic quark
model one has
$$h_1^q(x)=\Delta q(x)\eqno(29)$$
we will speculate that eq.~(29) holds for quarks and antiquarks.
The results of our calculations are shown in fig.~13 at two
different energies and we see that $A_{TT}$ has a similar trend
as $A_{LL}$ in figs.~12a, b, as a consequence of eq.~(29). Clearly
this prediction is only a guide for a future experiment at RHIC
which will indeed lead to the actual determination of $h_1(x)$.

\b

\item{\hbox to\parindent{\enskip {\bf 5.}\hfill}} {\bf CONCLUDING
REMARKS}

\m

We have seen that $W^{\pm}$ and $Z$ production in $pp$ and $pn$
collisions at RHIC up to $\sqrt s=500\ GeV$ will provide an
extremely valuable source of information on parton
distributions.  The high luminosity of the machine will allow
copious production of these gauge bosons and the precise
measurement of unpolarized cross sections will answer
unambigously the relevant question of flavor symmetry breaking
of the light sea quarks (or antiquarks) of the nucleon. Since
polarized proton beams will be available at RHIC with a high
polarization, it will be possible to use longitudinally polarized
beams to measure parity-violating asymmetries $A_L$. These
asymmetries are large and contain clear signatures of the
properties (magnitude and sign) of parton polarizations and it
will greatly improve the calibration of light quark and antiquark
polarized distributions. The absence of antiquark polarizations
implies a vanishing double helicity asymmetry $A_{LL}$, which is
rather easy to test. Of course, all these informations will be
complementary to the data extracted from polarized deep
inelastic scattering experiments both on proton and neutron
targets and will supply a detailed knowledge of the quark spin
structure of the nucleon. Finally the use of transversely
polarized proton beams for $Z$ production in $pp$ collisions will
allow the first determination of the transversity distribution
$h_1(x)$ for quarks and antiquarks, which is so far totally
unknown. To conclude, we hope the arguments given above are
strong enough to make gauge boson production a vital part of the
future experimental programme at RHIC with polarized proton
beams.

\vfill\eject

\n {\bf ACKNOWLEDGMENTS}

\m

One of us (J.~S.) acknowledges kind hospitality at the IFT
(Instituto de F\'\i sica Te\'orica) of S\~ao Paulo where part of
this work was done and is grateful to the Funda\c c\~ao de
Amparo \`a Pesquisa do Estado de S\~ao Paulo (FAPESP) for
providing financial support.

\vfill\eject

\n{\bf REFERENCES}

\m

\parindent 1truecm

\item{\hbox to\parindent{\enskip \hphantom{0}[1] \hfill}}
C.~Bourrely, J.Ph.~Guillet and J.~Soffer, Nucl. Phys. {\bf B
361} (1991) 72. Proceeding of the Polarized Collider Workshop,
University Park PA (1990), Eds J.~Collins, S.~Heppelmann and
R.W.~Robinett, AIP Conf. Proc. N$^{\circ}$ 223 AIP, New York
(1991).

\item{\hbox to\parindent{\enskip \hphantom{0}[2] \hfill}}
G.~Bunce et al. Particle World, vol.~3 (1992) 1.

\item{\hbox to\parindent{\enskip \hphantom{0}[3] \hfill}}
Proposal on Spin Physics using the RHIC Polarized Collider, R~5
(14 august 1992). Approved october 1993.

\item{\hbox to\parindent{\enskip \hphantom{0}[4] \hfill}}
P.~Chiappetta, P.~Colangelo, J.Ph.~Guillet and G.~Nardulli,
Z.~Phys. {\bf 59} (1993) 629.

\item{\hbox to\parindent{\enskip \hphantom{0}[5] \hfill}}
P.~Amaudruz et al. (New Muon Collaboration), Phys. Rev. Lett. {\bf
66} (1991) 2712~; D.~Allasia et al., Phys. Lett. {\bf B~249} (1990)
366.

\item{\hbox to\parindent{\enskip \hphantom{0}[6] \hfill}}
K.~Gottfried, Phys. Rev. Lett. {\bf 18} (1967) 1174.

\item{\hbox to\parindent{\enskip \hphantom{0}[7] \hfill}}
C.~Bourrely and J.~Soffer, Phys. Lett. {\bf B~314} (1993) 132.

\item{\hbox to\parindent{\enskip \hphantom{0}[8] \hfill}}
M.A.~Doncheski, F.~Halzen, C.S.~Kim and M.L.~Stong, Preprint
\hfill\break MAD/PH/744, june 1993.

\item{\hbox to\parindent{\enskip \hphantom{0}[9] \hfill}}
K.~Goulianos, private communication.

\item{\hbox to\parindent{\enskip [10] \hfill}} J.~Ralston and
D.E.~Soper, Nucl. Phys. {\bf B~152} (1979) 109~; J.L.~Cortes,
B.~Pire and J.P.~Ralston, Z.~Phys. {\bf C~55} (1992)
409~; R.~Jaffe and X.~Ji, Phys. Rev. Lett. {\bf 67} (1991)
552~; R.~Jaffe and X.~Ji, Nucl. Phys. {\bf B~375} (1992) 527~;
X.~Ji, Nucl. Phys. {\bf B~402} (1993) 217.

\item{\hbox to\parindent{\enskip [11] \hfill}} E.J.~Eichten,
I.~Hinchliffe and C.~Quigg, Phys. Rev. {\bf D~45} (1992) 2269.

\item{\hbox to\parindent{\enskip [12] \hfill}} F.~Buccella and
J.~Soffer, Mod. Phys. Lett. {\bf A~8} (1993) 225~; Europhysics
Letters {\bf 24} (1993) 165~; Phys. Rev. {\bf D~48} (1993) 5416.

\item{\hbox to\parindent{\enskip [13] \hfill}} C.~Bourrely et
al., Preprint CPT-93/P.2961 (october 1993) (to be pu\-bli\-shed in
Z.~Phys.).

\item{\hbox to\parindent{\enskip [14] \hfill}} J.~Ellis and
R.L.~Jaffe, Phys. Rev. {\bf D~9} (1974) 1444.

\item{\hbox to\parindent{\enskip [15] \hfill}} J.~Ashman et al.
(European Muon Collaboration), Phys. Lett. {\bf B~206} (1988)
364~; Nucl. Phys. {\bf B~328} (1989) 1.

\item{\hbox to\parindent{\enskip [16] \hfill}} P.L.~Anthony et al.
(E 142 Collaboration), Phys. Rev. Lett. {\bf 71} (1993) 959.

\vfill\eject

\item{\hbox to\parindent{\enskip [17] \hfill}} J.~Qiu and
G.~Sterman, Nucl. Phys. {\bf B~378} (1992) 52~; R.L.~Jaffe and
X.~Ji, Phys. Rev. {\bf D~43} (1991) 724.

\item{\hbox to\parindent{\enskip [18] \hfill}} X.~Ji, Phys.
Lett. {\bf B~234} (1992) 137~; R.L.~Jaffe and X.~Ji, Phys. Rev.
Lett. {\bf 71} (1993) 2547.

\vfill\eject

\n{\bf FIGURE CAPTIONS}

\m

\parindent 2truecm

\item{\hbox to\parindent{\enskip Fig. 1a \hfill}}
$d\sigma/dy$ versus $y$ for $W^+$ and $W^-$ production in $pp$
collisions at $\sqrt s=350\ GeV$.

\item{\hbox to\parindent{\enskip Fig. 1b \hfill}}
Same as (a) at $\sqrt s=500\ GeV$.

\item{\hbox to\parindent{\enskip Fig. 2 \hfill}}
$d\sigma/dy$ versus $y$ for $W^+$ production in $pn$ collisions at
$\sqrt s=350$ and $500\ GeV$. $W^-$ cross section
is obtained by symmetry around $y=0$.

\item{\hbox to\parindent{\enskip Fig. 3 \hfill}} The
ratio $R_W$ (see eq.~(7)) versus $y$ at $\sqrt s=350$ and
$500\ GeV$.

\item{\hbox to\parindent{\enskip Fig. 4 \hfill}}
$d\sigma/dy$ versus $y$ for $Z$ production in $pp$ collisions at
$\sqrt s=350$ and $500\ GeV$.

\item{\hbox to\parindent{\enskip Fig. 5 \hfill}}
$d\sigma/dy$ versus $y$ for $Z$ production in $pn$ collisions at
$\sqrt s=350$ and $500\ GeV$.

\item{\hbox to\parindent{\enskip Fig. 6a \hfill}} The
ratio $R_Z$ (see eq.~(10)) versus $y$ at $\sqrt s=350\ GeV$.
Dashed curve corresponds to $\delta\o q=0$, solid curve
corresponds to the choice in eq.~(6).

\item{\hbox to\parindent{\enskip Fig. 6b \hfill}}
Same as (a) at $\sqrt s=500\ GeV$.

\item{\hbox to\parindent{\enskip Fig. 7 \hfill}}
Parton polarizations $\Delta q/q$ versus $x$ taken at
$Q^2=M_W^2$ following eqs.~(15, 16a, 16b).

\item{\hbox to\parindent{\enskip Fig. 8a \hfill}} The
parity-violating helicity asymmetry $A_L$ versus
$y$ for $W^+$ and $W^-$ production in $pn$ collisions at $\sqrt
s=350\ GeV$.

\item{\hbox to\parindent{\enskip Fig. 8b \hfill}}
Same as (a) at $\sqrt s=500\ GeV$.

\item{\hbox to\parindent{\enskip Fig. 9a \hfill}} The
ratio $a_L^W$ (see eq.~(20)) versus $y$ at $\sqrt s=350\ GeV$.
Solid curve, $\delta\o q(x)\ne 0$ and $\Delta\o q(x)\ne 0$. Small
dashed curve, $\delta\o q(x)\ne 0$ and $\Delta\o q(x)=0$.

\item{\hbox to\parindent{\enskip Fig. 9b \hfill}}
Same as (a) at $\sqrt s=500\ GeV$.

\item{\hbox to\parindent{\enskip Fig. 10 \hfill}} The
parity-violating helicity asymmetry $A_L$ versus
$y$ for $Z$ production in $pn$ collisions at $\sqrt s=350$ and
$500\ GeV$.

\item{\hbox to\parindent{\enskip Fig. 11a \hfill}} The ratio
$a_L^Z$ (see eq.~(22)) versus $y$ at $\sqrt s=350\ GeV$. Solid
curve, $\delta\o q(x)\ne 0$ and $\Delta\o q(x)\ne 0$. Small
dashed curve, $\delta\o q(x)\ne 0$ and $\Delta\o q(x)=0$.

\item{\hbox to\parindent{\enskip Fig. 11b \hfill}} Same as (a) at
$\sqrt s=500\ GeV$.

\item{\hbox to\parindent{\enskip Fig. 12a \hfill}}
Parity-conserving double helicity asymmetry $A_{LL}$ versus
$y$ for $W^{\pm}$ and $Z$ production in $pp$ collisions at $\sqrt
s=350\ GeV$.

\item{\hbox to\parindent{\enskip Fig. 12b \hfill}} Same as (a) at
$\sqrt s=500\ GeV$.

\item{\hbox to\parindent{\enskip Fig. 13 \hfill}} Double spin
transverse asymmetry $A_{TT}$ versus $y$ for $Z$
production in $pp$ collisions at $\sqrt s=350$ and $500\ GeV$.

 \end